\documentstyle[prd,aps,twocolumn,epsf,axodraw]{revtex}
\def\br{\begin{eqnarray}}
\def\er{\end{eqnarray}}
\def\be{\begin{equation}}
\def\ee{\end{equation}}

\def\({\left(}
\def\){\right)}
\def\s{\sigma}

\begin{document}
\twocolumn[\hsize\textwidth\columnwidth\hsize\csname 
@twocolumnfalse\endcsname                            
%
\title{
Two-photon final states in peripheral heavy ion collisions }
\author{A.~A.~Natale, C.~G.~Rold\~ao and J.~P.~V.~Carneiro \\
}
\address{
Instituto de F\'{\i}sica Te\'orica\\
Universidade Estadual Paulista\\
Rua Pamplona 145\\
01405-900, S\~ao Paulo, SP\\
Brazil}
\date{\today}
\maketitle
\begin{abstract}

We discuss processes leading to two photon final states in
peripheral heavy ion collisions at RHIC. Due to the large photon
luminosity we show that the continuum subprocess $\gamma \gamma
\rightarrow \gamma \gamma$ can be observed with a large number of
events. We study this reaction when it is intermediated by a
resonance made of quarks or gluons and discuss its interplay with
the continuum process, verifying that in several cases the
resonant process ovewhelms the continuum one. It is also
investigated the possibility of observing a scalar resonance (the
$\sigma$ meson) in this process. Assuming for the $\sigma$ the
mass and total decay width values recently reported by the E791
Collaboration we show that RHIC may detect this particle in its
two photon decay mode if its partial photonic decay width is of
the order of the ones discussed in the literature.
\end{abstract}
\vskip 0.5cm]                           

\section{Introduction}
In the Relativistic Heavy-Ion Collider (RHIC) operating at the
Brookhaven Laboratory beams of heavy-ions are colliding with the
main interest in the search of a quark-gluon plasma in central
nuclear reactions. In addition to this important feature of
heavy-ion colliders, peripheral collisions may give rise to a huge
luminosity of photons opening the possibilities of studying
two-photon and other interactions as discussed by several authors
\cite{baur}.

In order to study peripheral $\gamma - \gamma$ heavy-ion
collisions we have to pay attention to two important facts, which
are fundamental to determine the kind of physics that is truly
accessible in these interactions. The first one is to avoid
events where hadronic particle production overshadows the $\gamma
- \gamma $ interaction, i.e., events where the nuclei physically
collide (with impact parameter $b < 2 R_A$, $R_A$ being the
nuclear radius) are excluded from calculations of the usable
luminosity.This exclusion has been discussed by Baur\cite{gbaur}
and can be handled as described by Cahn and Jackson \cite{cahn}.
The second point to remember is that the photons will carry only
a small fraction of the ion momentum, favoring low mass final
states. Therefore, hadronic resonances with masses up to a few
GeV can be produced at large rates \cite{bertulani,natale}, as
well as any two-photon process not leading to extremely heavy
final states.

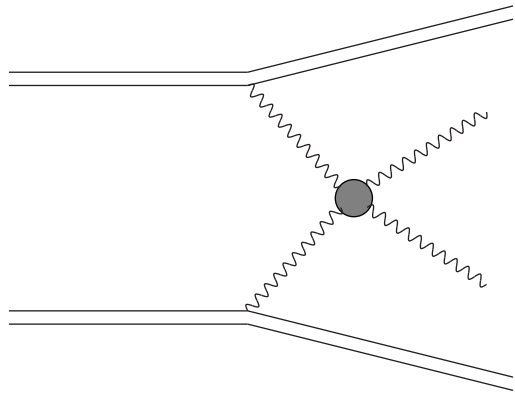
\begin{figure}[htb]
\vskip 1in
\begin{picture}(290,10)(10,210)
 \Line(30,255)(120,255)
 \Line(30,250)(120,250)
 \Line(30,165)(120,165)
 \Line(30,160)(120,160)
 \GCirc(160,207.5){7}{0.5}
 \Photon(120,250)(155,212){2}{10}
 \Photon(120,165)(155,204){2}{10}
 \Photon(165,212)(210,240){2}{10}
 \Photon(165,204)(210,175){2}{10}
 \Line(120,255)(220,280)
 \Line(120,250)(220,275)
 \Line(120,165)(220,140)
 \Line(120,160)(220,135)
\end{picture}
\label{diagrama}
 \vskip 1.0in
  \caption{ Diagram for $\gamma \gamma $ fusion in a peripheral heavy-ion collision.
 The blob represents a continuum or resonant process leading to a two-photon final
 state.  }
\end{figure}

One of the purposes of our work was  motivated by the possibility
of measurement of the continuum process $\gamma \gamma \rightarrow
\gamma \gamma$ , see Fig.(\ref{diagrama}), as suggested in Ref.
\cite{bertulani}, as well as the resonant one $\gamma \gamma
\rightarrow R \rightarrow \gamma \gamma $ which was predicted in
Ref. \cite{natale} as a clear signal for resonances made of quarks
or gluons. The continuum $\gamma \gamma \rightarrow \gamma
\gamma$ reaction is interesting per se but is also a background
for the resonant process. This last one is quite important to be
observed because it involves only the electromagnetic coupling of
the resonance. Its knowledge with high precision is very useful,
for instance, to unravel the possible amount of mixing in some
glueball candidates \cite{close}, complementing the information
obtained through the observation of hadronic decays. Another
interesting study is the possible production of a light scalar
meson ($\sigma$).

The E-791 experiment recently reported that the $\sigma $ meson
has been observed in its $\pi \pi$ decay mode \cite{e791}, which
is the same final state questioned over many years as a possible
(or not) signal for this broad resonance. We discuss the
production of this elusive meson in peripheral heavy-ion
collisions. However, we focus attention on the observation of its
$\gamma \gamma $ decay. We believe that the observation of this
decay could settle definitively the existence of this particle as
well as of its quark (or gluon) content. Although its branching
ratio into photons is quite small compared to the main one ($\pi
\pi$), we will argue that the high photon luminosity in peripheral
heavy-ion  collisions may allow its observation, as long as it
has the very same properties predicted in Ref. \cite{e791}.

Photons with invariant mass of O($1$ \, GeV) are also one
possible signal for the quark-gluon plasma, therefore, the
detectors at RHIC have high efficiency to detect these photons,
which is a welcomed capability in regard to the final states that
we discuss in this work. The background for the two-photon
initiated process were discussed by Nystrand and Klein
\cite{klein}. Double-Pomeron exchange may be a background source
\cite{engel}. However, they are negligible after imposing the
peripheral condition $b > 2R_A$ as long as we operate with very
heavy ions\cite{eu}. The cuts to remove the background are
discussed in Ref. \cite{klein}, i.e., the final state must have:
a) low multiplicity, b) small summed transverse momentum and c)
be centered around rapidity 0 with a fairly narrow width of
center-of-mass rapidity ($y_{cm} \lesssim 1 - 2$). These cuts are
easily implemented when we analyze the reaction proposed in this
work: $ZZ \rightarrow ZZ \gamma \gamma$.

The remainder of the paper is organized as follows. In the second
section we introduce the photon distribution in the ion that will
be used in our calculations. Section III contains the discussion
of the $\gamma \gamma \rightarrow \gamma \gamma $ process and the
cuts that we will adopt for this reaction.  In section IV we
present values for the production of resonances made of quarks
(or gluons), and section V contains a discussion about the
possibility of observing the $\sigma $ meson in its two photon
decay. The last section contains our conclusions.

\section{Photon distribution in the Ion}

The photon distribution in the nucleus can be described using the
equivalent-photon or Weizs\"{a}cker-Williams approximation in the
impact parameter space. Denoting by $F(x)dx$ the number of
photons carrying a fraction between $x$ and $x+dx$ of the total
momentum of a nucleus of charge $Ze$, we can define the
two-photon luminosity through
\begin{equation}
\frac{dL}{d\tau} = \int ^1 _\tau \frac{dx}{x} F(x) F(\tau/x),
\end{equation}

\noindent
 where $\tau = {\hat s}/s$, $\hat s$ is the square of
the center of mass (c.m.s.) system energy of the two photons and
$s$ of the ion-ion system. The total cross section of the process
$ ZZ \rightarrow ZZ \gamma \gamma $ is
\begin{equation}
\sigma (s) = \int d\tau \frac{dL}{d\tau} \hat \sigma(\hat s),
\label{sigfoton}
\end{equation}

\noindent
 where $ \hat \sigma(\hat s)$ is the cross-section of
the subprocess $\gamma \gamma \rightarrow \gamma \gamma$.

There remains only to determine $F(x)$. In the literature there
are several approaches for doing so, and we choose the
conservative and more realistic photon distribution of
Ref.\cite{cahn}. Cahn and Jackson~\cite{cahn}, using a
prescription proposed by Baur~\cite{gbaur}, obtained a photon
distribution which is not factorizable. However, they were able
to give a fit for the differential luminosity which is quite
useful in practical calculations:
\begin{equation}
\frac{dL}{d\tau}=\left(\frac{Z^2 \alpha}{\pi}\right)^2
\frac{16}{3\tau} \xi (z),
 \label{dl1}
\end{equation}

\noindent
 where $z=2MR\sqrt{\tau}$, $M$ is the nucleus mass, $R$
its radius and $\xi(z)$ is given by
\begin{equation}
\xi(z)=\sum_{i=1}^{3} A_{i} e^{-b_{i}z},
 \label{dl2}
\end{equation}
which is a fit resulting from the numerical integration of the
photon distribution, accurate to $2\% $ or better for
$0.05<z<5.0$, and where $A_{1}=1.909$, $A_{2}=12.35$,
$A_{3}=46.28$, $b_{1}=2.566$, $b_{2}=4.948$, and $b_{3}=15.21$.
For $z<0.05$ we use the expression (see Ref.~\cite{cahn})
\begin{equation}
\frac{dL}{d\tau}=\left(\frac{Z^2 \alpha}{\pi}\right)^2
\frac{16}{3\tau}\left(\ln{(\frac{1.234}{z})}\right)^3 .
 \label{e5}
\end{equation}
The condition for realistic peripheral collisions ($b_{min} > R_1
+ R_2$) is present in the photon distributions showed above.

The two photon production in heavy ions collisions can also be
intermediated by the diffractive subprocess $P P \rightarrow
\gamma \gamma $, where $P$ is the Pomeron. But, as was shown by
some of us in Ref.\cite{eu}, when the cut in the impact parameter
is introduced the double-Pomerom exchange is not important for
very heavy nuclei.

\section{The continuum reaction $\gamma + \gamma \rightarrow \gamma + \gamma $}

The reaction $\gamma \gamma \rightarrow \gamma \gamma $ (given by
the fermion box diagram) was first calculated exactly by Karplus
and Neuman using the usual Feynman techniques \cite{karplus}.
Some time later this process was calculated by De Tollis using
the Mandelstam representation and dispersion
relations\cite{tollis2}. They encountered relatively simple
helicity amplitudes for this process. Of the sixteen helicity
amplitudes, due to symmetry properties, the number of independent
amplitudes will be only five \cite{landau,tollis1}, that may be
chosen to be $M_{++++}$, $M_{++--}$, $M_{+-+-}$, $M_{+--+}$ and
$M_{+++-}$. Where the $+$ or $-$ denotes the circular
polarization values $+1$ and $-1$. The remaining helicity
amplitudes may be obtained from parity and permutation symmetry.

Of these five helicity amplitudes, three are related by crossing,
hence it is sufficient to give just three, that in agreement with
Ref. \cite{tollis2,tollis1} are
\begin{eqnarray}
M_{++++} &=& 1 + \left\{ 2 +\frac{4s_t}{r_t} \right\} B(s_t)
 \nonumber \\
 &+&
\left\{ 2 +\frac{4t_t}{r_t} \right\} B(t_t)
\nonumber \\
&+& \left\{\frac{2(s_t^2 + t_t^2)}{r_t^2} - \frac{2}{r_t} \right\}
[T(s_t) + T(t_t)]
\nonumber \\
 &+& \left\{ -\frac{1}{s_t} +
\frac{1}{2 r_t s_t}\right\} I(r_t,s_t)
 \nonumber \\
 &+& \left\{ -\frac{1}{t_t} +
\frac{1}{2 r_t t_t}\right\} I(r_t,t_t)
\nonumber \\
&+& \left\{ -\frac{2(s_t^2 + t_t^2)}{r_t^2} + \frac{4}{r_t} +
\frac{1}{s_t} \right.
 \nonumber \\
 &+& \left. \frac{1}{t_t} + \frac{1}{2 s_t t_t}
\right\} I(s_t,t_t),
\nonumber \\
M_{+++-} &=& -1 + \left\{ -\frac{1}{r_t} - \frac{1}{s_t} -
\frac{1}{t_t} \right\}
 \nonumber \\
 &\times & [T(r_t) + T(s_t) + T(t_t)]
\nonumber \\
&+& \left\{ \frac{1}{t_t} + \frac{1}{2 r_t s_t} \right\}
I(r_t,s_t)
 \nonumber \\
 &+& \left\{ \frac{1}{s_t} + \frac{1}{2 r_t s_t} \right\}
I(r_t,s_t)
\nonumber \\
&+& \left\{ \frac{1}{r_t} + \frac{1}{2 s_t t_t} \right\}
I(s_t,t_t), \nonumber \\
M_{++--} &=& -1 + \frac{1}{2 r_t s_t} I(r_t,s_t)
 \nonumber \\
 &+& \frac{1}{2 r_t
st_t} I(r_t,t_t) + \frac{1}{2 s_t t_t} I(s_t,t_t),
\label{elemento}
\end{eqnarray}

\noindent
 where $r_t$, $s_t$ and $t_t$ are related with the standard
 Mandelstam variables s, t, and u by
\begin{eqnarray}
r_t = \frac{1}{4} \frac{s}{m_f^2}, \,\,\,\,\, s_t = \frac{1}{4}
\frac{t}{m_f^2}, \,\,\,\,\, t_t = \frac{1}{4} \frac{u}{m_f^2},
\nonumber
\end{eqnarray}

\noindent
 $m_f$ is the mass of the fermion in the loop.

The remaining thirteen amplitudes may be obtained from these by
using the relations
\begin{eqnarray}
M_{+++-} = M_{++-+} &=& M_{+-++} = M_{-+++} \nonumber \\
M_{+-+-}(s,t,u) &=& M_{++++}(u,t,s) \nonumber \\
M_{+--+}(s,t,u) &=& M_{++++}(t,s,u) \nonumber \\
\end{eqnarray}

\noindent
 where $s$, $t$ and $u$ are the Mandelstam variables.

The transcendental functions $B$, $T$ and $I$ that appear in
 Eq.(\ref{elemento}) were defined by Karplus and Neuman
 \cite{karplus}, and are
\begin{eqnarray}
B(x) &=& \sqrt{1 - \frac{1}{x}} \mbox{arcsinh}(\sqrt{-x}) -1,
\,\,\,\,\,\,\,\,\,\,\,\,\,\,\,\,\,\,\,\,\,\,\,\,\,\,\,\,\,\,\,\,\,  (x<0) \nonumber \\
&=& \sqrt{1 - \frac{1}{x}} \arcsin(\sqrt{x}) -1;,
\,\,\,\,\,\,\,\,\,\,\,\,\,\,\,\,\,\,\,\,\,\,\,\,\,\,\,\,\,\,\,\,\,\,\,\,\,
(0<x<1) \nonumber \\
&=& \sqrt{1 - \frac{1}{x}} \mbox{arccosh}(\sqrt{x}) - 1 -
\frac{\pi i}{2} \sqrt{1 - \frac{1}{x}}, \,\,\, (1<x) \label{beta}
\end{eqnarray}
\begin{eqnarray}
T(x) &=& [\mbox{arcsinh}(\sqrt{-x})]^2,
\,\,\,\,\,\,\,\,\,\,\,\,\,\,\,\,\,\,\,\,\,\,\,\,\,\,\,\,\,\,\,\,\,\,\,\,\,\,\,\,
\,\,\,\,\,\,\,\,\,\,\,\,\,\,\,
(x<0) \nonumber \\
&=& - [\arcsin(\sqrt{x})]^2,
\,\,\,\,\,\,\,\,\,\,\,\,\,\,\,\,\,\,\,\,\,\,\,\,\,\,\,\,\,\,\,\,\,\,\,\,\,\,\,\,
\,\,\,\,\,\,\,\,\,\,\,\,\,\,\,\,\,\,
(0<x<1) \nonumber \\
&=& [\mbox{arccosh}(\sqrt{x})]^2 - \frac{\pi^2}{4} -i \pi
\mbox{arcosh}(\sqrt{x}), \,\, (1<x) \label{teta}
\end{eqnarray}
\begin{eqnarray}
I(x,y) &=& I(y,x), \nonumber \\
Re\{ I(x,y) \} &=& \frac{1}{2a}Re\left\{ \Phi \left(
\frac{a+1}{a+b(x)} \right) +  \Phi \left( \frac{a+1}{a-b(x)}
\right) \right. \nonumber \\
&-& \left. \Phi \left( \frac{a-1}{a+b(x)} \right) - \Phi \left(
\frac{a-1}{a-b(x)} \right) + (x \leftrightarrow y) \right\}
\nonumber \\
Im\{ I(x,y) \} &=& - \frac{\pi}{2a} \ln{y(a + b(x))^2}, \,\,\,\,
(x \geq 1) \nonumber \\
&=& - \frac{\pi}{2a} \ln{x(a + b(y))^2}, \,\,\,\, (y \geq 1),
\label{ipsilon}
\end{eqnarray}

\noindent
 where $a = \sqrt{1 -(x+y)/xy}$, $x$, $y$ = $r_t$, $s_t$
and $t_t$, and
\begin{eqnarray}
b(x) &=& \sqrt{1 - \left( \frac{1}{x} \right)}, \,\,\,\,\,\,
\mbox{when} \,\,\,\, (x<0) \,\,\,\, \mbox{or} \,\,\,\, (x>1)
\nonumber \\
&=& i \sqrt{1 - \left( \frac{1}{x} \right)}, \,\,\,\, \mbox{when}
\,\,\,\, (0<x<1).
 \label{b}
\end{eqnarray}

The Spence function, $\Phi(z)$, is defined as $ \Phi(z) = \int ^z
_0 \ln(1-t)dt/t$, and its properties can been found in Ref.
\cite{spence}.

The differential cross section of photon pair production from
photon fusion, i.e. the box diagram, is
\begin{equation}
\frac{d \sigma}{d \cos \theta }= \frac{1}{2 \pi} \frac{\alpha
^4}{s} (\sum_f q^2_f)^4 \sum |M|^2.
 \label{dcaixa}
\end{equation}

\noindent
  $\theta $ is the scattering angle,  $\alpha $ is the fine-structure
 constant, $q_f$ is the charge
 of the fermion in the loop and the sum is over the leptons $e$ and
 $\mu $ and the quarks $u$, $d$ and $s$, which are the relevant
 particles in the mass scale that we shall discuss. Another possible
 contribution to this continuum process comes from pion loops, which, apart
 from possible double counting, were shown to be negligible compared
 to the above one\cite{tollis3}.
The second sum is over the
 sixteen helicity  amplitudes, $M_{\lambda _1 \lambda_2 \lambda _3
 \lambda_4}$, where $\lambda _1$ and $\lambda _2$ correspond to
  polarizations of incoming photons and  $\lambda _3$ and $\lambda _4$
  for the outgoing photons.
The matrix elements sumed over the final polarizations and
averaged over the initial polarizations is given by
\begin{eqnarray}
\sum |M|^2 &=& \frac{1}{2} \{ |M_{++++}|^2 + |M_{++--}|^2 +
|M_{+-+-}|^2  \nonumber \\
&+&  |M_{+--+}|^2 + 4 |M_{+++-}|^2 \}.
 \nonumber
\end{eqnarray}

We consider the scattering of light by light, that is, the
reaction $\gamma \gamma \rightarrow \gamma \gamma$, in Au-Au
collisions for energies available at RHIC, $\sqrt{s} = 200$
GeV/nucleon.  We checked our numerical code reproducing the many
results of the literature for the box subprocess, including
asymptotic expressions for the low and high energies compared to
the fermion mass present in the loop, and the peak value of the
cross section (see, for instance, Ref.\cite{landau}).

\begin{figure}[htb]
\epsfxsize=.45\textwidth
\begin{center}
\leavevmode
 \epsfbox{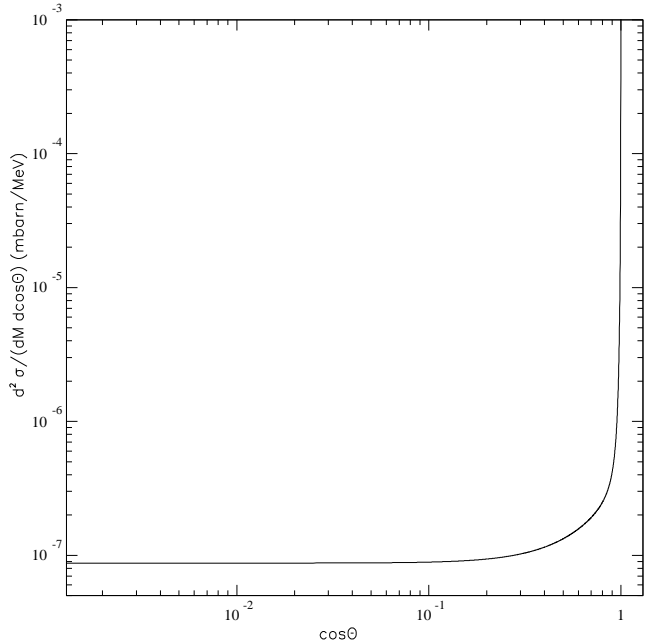}
\end{center}
\vskip -1.6cm
 \caption{Angular distribution of $ZZ \rightarrow ZZ\gamma\gamma$ scattering
 at an invariant mass of 500 MeV. The scattering  angle
 $\theta $ is in the photon pair center-of-mass system.}
  \label{dsigma_dcos}
\end{figure}

 In Fig.(\ref{dsigma_dcos}) the dependence of the ion-ion cross
section with the cosine of the scattering angle $\theta $, in the
two photon center-of-mass system, is shown for an invariant photon
pair mass equal to 500 MeV. It is possible to observe that the
cross section is strongly peaked in the backward direction, but is
relatively flat out to $\cos{\theta} \approx 0.4$, where it starts
rising very fast. It is symmetric with respect to $\theta $ and
the same behavior is present in the forward direction.

\begin{figure}[htb]
\epsfxsize=.45\textwidth
\begin{center}
\leavevmode
 \epsfbox{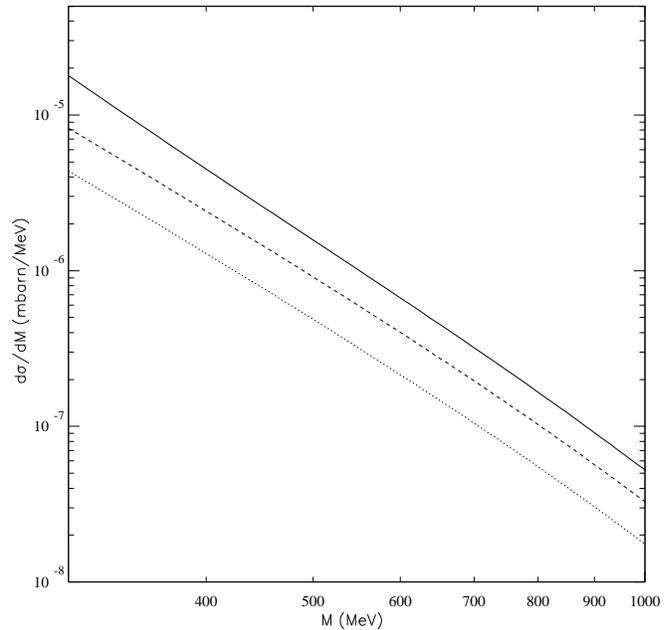}
\end{center}
\vskip -1.6cm
 \caption{ Invariant mass distribution of photon pair
 production at RIHC via photon fusion through a
 fermion loop. Each curve represents a different cut in
 the scattering angle, $\theta $, the solid line
 is for $10 ^o < \theta < 170^o$, the dashed is for
 $40^o < \theta < 140^o$,  and the dotted one for
 $60^o < \theta < 120^o$. }
  \label{caixa_teta}
\end{figure}

  The photon pair production in ion collisions over an invariant
mass range between 300 and 1000 MeV for different angular cuts is
shown in Fig.(\ref{caixa_teta}). It is possible to observe that
the cross sections is strongly restricted when the angular cut is
more drastic, the difference between the cuts $10^o < \theta
<170^o$ and $60^o < \theta < 120^o$ is quite large, the cross
section is strongly suppressed for the second range. We will
impose in all the calculations throughout this work a cut in the
scattering angle equal to $|\cos \theta | = 0.5$. This cut is
conservative, but it will make possible to compare the cross
section of the box diagram with rival processes, that will be
discussed in the following sections, as well as it is enough to
eliminate the effect of double bremsstrahlung (which dominates the
region of $| \cos{\theta} | \approx 1$). Finally, this kind of cut
is totally consistent with the requirements proposed in
Ref.\cite{klein}

 \vskip -1.3 cm

The fermions that contribute in the box diagram are the leptons
$e$ and $\mu$ and the quarks $u$, $d$ and $s$, which are
important for the mass range that we are interested (heavier
quarks will give insignificant contributions and the same is true
for the charged weak bosons). We assumed for their masses the
following values: $m_e = 0.5109$ MeV, $m_\mu = 105.6584$ MeV,
$m_u = 5$ MeV, $m_d = 9$ MeV and $m_s = 170$ MeV.
 \vskip -.5in
\vskip -1.7cm
\begin{figure}[htb]
\epsfxsize=.45\textwidth
\begin{center}
\leavevmode
 \epsfbox{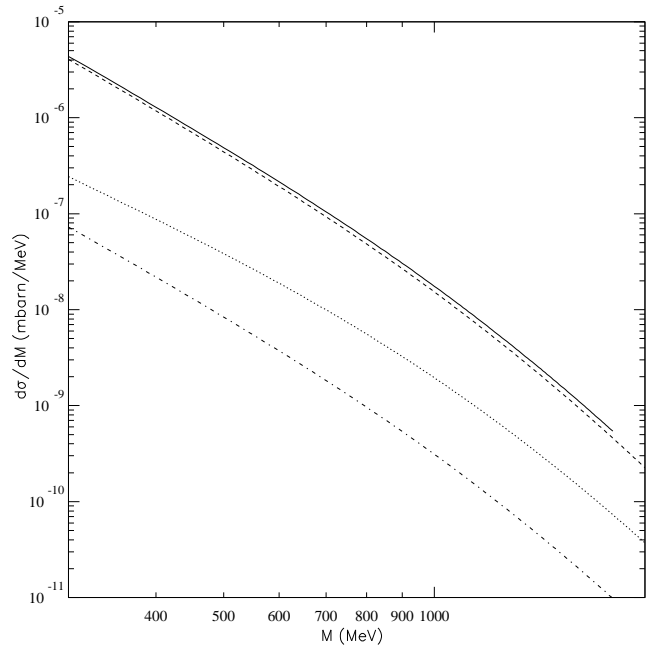}
\end{center}
\vskip -1.6cm
 \caption{ Invariant mass distribution of photon pair production
 at RHIC center of mass energy of $\sqrt{s} = 200$ GeV for each fermion loop.
The solid line denotes the differential cross section for all
fermions,
 the dashed one is the electron loop contribution, the doted is the muon one, the
 dash-dotted is the $u$ quark contribution. The result is obtained
 with the cut $|\cos{\theta}| <0.5$.}
  \label{fermion_loop}
\end{figure}
\vskip .7cm

The electron gives the major contribution to the total result as
can be seen in Fig.(\ref{fermion_loop}). The second most important
 contribution is due to the muon, but it is at least one order of
 magnitude smaller than the electron one.
The $d$ and $s$ quark contribution (up to O(2 GeV)) are smaller
due to their masses and charges, because the process is
proportional to $(q_f ^2)^4$  where $q_f$ is their charge. Their
contribution is also insignificant compared to the electron
result.

As discussed in Ref.\cite{bertulani} the $\gamma \gamma$
scattering can indeed be measured in peripheral heavy ion
collisions. The cut in the angular distribution gives back to
back photons in the central rapidity region free of the
background. However, as we shall see in the next section, there
are gaps where the $\gamma \gamma \rightarrow \gamma \gamma$
process is overwhelmed by the presence of resonances like the
$\eta$, $\eta\prime$ and others. Even the broad $\s$ resonance
could be of the order of the continuum process. Just to give one
idea of the number of events, with a luminosity of 2.0 $\times
10^{26}$ cm$^{-2}$s$^{-1}$ \cite{klein} and choosing a bin of
energy of 200 MeV, centered at the energy of 700 MeV (which is
free of any strong resonance decaying into two-photons), we have
1532 events/year assuming $100\%$ efficiency in the tagging of
the ions and photon detection.

\section{Resonances decaying into two-photons}

Photon pair production via the box diagram is a background to
$\gamma \gamma \rightarrow R \rightarrow \gamma \gamma $ process
(or vice versa), both have the same initial and final states, and
for this reason they can interfere one in another. Normally the
interference between a resonance and a continuum process is
unimportant, because on resonance the two are out of phase.

The total cross section for the elementary subprocess $\gamma
\gamma \rightarrow R \rightarrow \gamma \gamma $ assuming a
Breit-Wigner profile is
\begin{eqnarray}
\frac{d \sigma ^{\gamma \gamma}_{ZZ}}{dM} = 16 \pi \frac{dL}{dM}
\frac{\Gamma ^2 (R \rightarrow \gamma \gamma)} {(M^2-m_R^2)^2
+m_R^2 \Gamma^2_{total}},
 \label{dsigfoton}
\end{eqnarray}

 \noindent
 where $M$ is the energy of the photons created by
the collision of the ions. $\Gamma (R\rightarrow \gamma \gamma
)$($\equiv \Gamma_{\gamma \gamma}$) and $\Gamma_{total}$ are the
partial and total decay width of the
 resonance with mass $m_R$ in its rest frame.

\begin{figure}[htb]
\epsfxsize=.45\textwidth
\begin{center}
\leavevmode \epsfbox{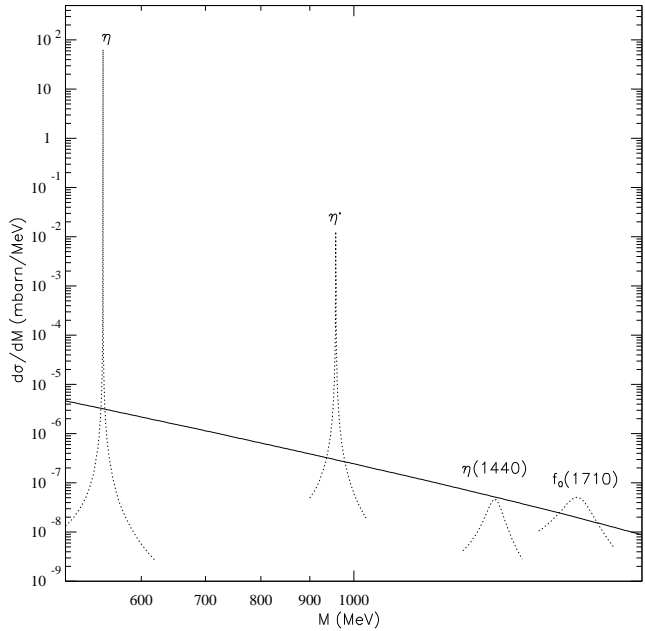}
\end{center}
\vskip -1.6cm
 \caption{Invariant mass distribution of photon production (with
the cut $|\cos{\theta}|<0.5$). The solid curve is for
 the box diagram, the dashed curves are due to the process
 $\gamma \gamma \rightarrow R \rightarrow \gamma \gamma $, where $R$ are
 the pseudoscalars resonances $\eta$ and $\eta^\prime$ and the glueballs
candidates $\eta(1410)$ and $f_0(1710)$.  }
  \label{glueball}
\end{figure}

 We are going to discuss only $J=0$ resonances made of quarks as
 well of gluons. The reaction $\gamma \gamma \rightarrow \pi^0
 \rightarrow \gamma \gamma$ was already discussed many years ago\cite{ora},
 where it was claimed that the interference vanishes. This result was
 criticized by De Tollis and Violini~\cite{tollis3}, affirming (correctly)
 that the interference exists. However, as we will discuss afterwards,
off resonance the interference is negligible. If the interference
is neglected, Eq.(\ref{dsigfoton}) can be used and we show in
Fig.(\ref{glueball}) the result for some resonance production
($\eta, \, \eta^\prime, \, \eta(1440), \, f_0(1710)$), whose
invariant mass of the produced photon pair is between 500 MeV and
2000 MeV. For comparison we also show the curve of the continuum
process. It is possible to see in that figure the well pronounced
peaks of the resonances $\eta$ and $\eta ^\prime$. We assumed for
their masses the values of 547.3 MeV and 957.78 MeV,
respectively, the $\eta$ total decay width is equal to 1.18 KeV
and the $\eta^\prime$ one is equal to 0.203 MeV. Their partial
decay width into photons are 0.46 KeV ($\eta$) and 4.06 KeV
($\eta^\prime$).

In the same figure we can see the predicted cross section for the
glueball candidates production in peripheral heavy ion collisions
by double photon fusion.

 We restrict the analysis to the $J=0$
glueballs candidates $\eta(1440)$ and $f_0(1710)$. For the
$\eta(1440)$ we used the mass and total decay width values of
Ref.\cite{pdg}, $m_R = 1405$ MeV, $\Gamma_{total} = 56$ MeV, for
the decay width into photons we use the value given in
Ref.\cite{close2}, $\Gamma_{\gamma \gamma} = 5.4$ keV. We see in
Fig.(\ref{glueball}) that the peak for this resonance is of the
same order of the continuum process. For the other glueball
candidate, $f_0(1710)$, the peak is clearly above the background.
For this one we assumed the values listed in Ref.\cite{pdg} of
mass and total width, $m_R = 1715$ MeV  $\Gamma_{total} = 125$
MeV, and for the two-photon decay width we adopted the value
encountered by the ALEPH Collab. \cite{aleph}, $\Gamma_{\gamma
\gamma} = 21.25$ keV. In all these cases the resonances can be
easily studied in peripheral heavy ion collisions.

Off resonance we can expect a negligible contribution for the
process $\gamma \gamma \rightarrow R \rightarrow \gamma \gamma $
and consequently the same for its interference with the continuum
process. However, it is instructive to present a more detailed
argument about why the interference can be neglected. In order to
do so we are obliged to introduce a model to calculate the
amplitudes for the process $\gamma \gamma \rightarrow R
\rightarrow \gamma \gamma $. These amplitudes will be computed
with the help of an effective Lagrangian for the pseudoscalar
interaction with photons (the scalar case will be discussed in
the next section), which is given by $g_p \varepsilon_{\lambda
\mu \nu \rho} F^{\lambda \mu} F^{\nu \rho} \Phi_p $, where $g_p$
is the coupling of the photons to the pseudoscalar field
$\Phi_p$, $\varepsilon_{\lambda \mu \nu \rho}$ is the
antisymmetric tensor and $F^{\lambda \mu}$ is the electromagnetic
field four-tensor.

The amplitudes for $\gamma \gamma \rightarrow \gamma \gamma $
intermediated by a pseudoscalar hadronic resonance are
~\cite{tollis3}:
\begin{eqnarray}
M_{++++} &=& \frac{2 \pi}{\alpha^2} F(\lambda r_t),
 \nonumber \\
M_{+-+-} &=& \frac{2 \pi}{\alpha^2} F(\lambda t_t),
 \nonumber \\
M_{+--+} &=& \frac{2 \pi}{\alpha^2} F(\lambda s_t),
 \nonumber \\
 M_{+++-} &=& 0,
  \nonumber \\
 M_{++--} &=& - \frac{2 \pi}{\alpha ^2} \{ F(\lambda s_t) + F(\lambda
 t_t) + F(\lambda r_t) \} ,
\label{pseudoescalar}
\end{eqnarray}

\noindent
 where $\alpha $ is the fine-structure constant, $\lambda
= (m_f/m_R)^2$, and
\begin{eqnarray}
F(x) = 16 x^2\frac{\Gamma_{\gamma \gamma}}{m_R} \left( 4 x - 1 + i
\frac{\Gamma _{total}}{m_R} \right)^{-1}.
 \label{fun_res}
\end{eqnarray}

The presence of the fine-structure constant in
 Eq.(\ref{pseudoescalar}) is a consequence of the fact that the
 amplitudes $M$ in these equations will be used in
 Eq.(\ref{dcaixa}), so it is necessary to get the correct
 dependence of the partial cross section with this constant.

A numerical evaluation of the cross section using
Eq.(\ref{pseudoescalar}) (for the same resonances present in
Fig.(\ref{glueball})) shows a totally negligible effect off
resonance in comparison with the box contribution. On resonance
the two processes are out of phase and the interference is
absent. We can now proceed with an argument showing that the
interference is not important. Let us assume that off resonance
the processes are in phase, and for a moment we forget the $t$
and $u$ channels contribution in Eq.(\ref{pseudoescalar}). The
$s$ channel contribution can be written as $M_2/(s - m_R^2 + i
\Gamma_R m_R)$, and denoting the continuum contribution by $M_1$
we can write the following interference term
\begin{eqnarray}
& & 2 \frac{s}{(s - m^2_R)^2 + \Gamma^2_R m^2_R} [ ( Re M_ 1 Re
M_2 + Im M_1 Im M_2 )
 \nonumber \\
  & & (s - m^2_R) + ( Re M_1 Im M_ 2 - Im M_1 Re M_ 2
) \Gamma_R m_R ].
 \nonumber
\end{eqnarray}

\noindent
 We can verify that the term proportional to $ s -
m^2_R$ integrates to zero when integrated in a bin centered at
$m^2_R$. With the second term the situation is different. If $Im
M_1$ or $Im M_2 \neq 0$ (assuming $ ReM_1 $ and $Re M_2 \neq 0$)
then there is a nontrivial interference. However, since we are
dealing with $J=0$ amplitudes, the only nonvanishing helicity
amplitudes are those in which the initial helicities and the
final helicities are equal. Inspection of the $M_{++++}$ and
$M_{++--}$ amplitudes of the box diagram (in the limit $m_f \ll
m_R$) shows that they are purely real, and the same happens with
$M_2$ (obtained from the $s$ channel contribution of
Eq.(\ref{pseudoescalar})), resulting in a vanishing interference!
Of course, this analysis is model dependent. In particular, at
the quark level the coupling $g_p$ has to be substituted by a
triangle diagram, which may have a real as well as an imaginary
part (see, for instance, Ref.\cite{novaes}). However, this
coupling is real for heavy quarks and its imaginary part is quite
suppressed if the resonance couples mostly to light quarks ($m_f
\ll m_R$, what happens for the resonances that we are considering
in the case of the u and d quarks, for the s quark the suppresion
is not so strong in the $\eta$ case, but we still have an extra
suppression due to its electric charge). Finally, the
interference does appear when we consider the amplitudes with the
resonance exchanged in the $t$ and $u$ channels, but it is easy
to see that they are kinematically suppressed and also
proportional to the small value of the total decay width. If the
total width is large (as we shall discuss in the next section)
the interference cannot be neglected. Although the above argument
has to rely on models for the low energy hadronic physics, we
believe that the direct comparison between the resonant and the
continuum processes, as presented in Fig.(\ref{glueball}), is
fairly representative of the actual result.

In Table I we show the number of events above background for the
$\eta$, $\eta^\prime$ and  $f_0(1710)$ which, as shown in
Fig.(\ref{glueball}) are clearly above the box contribution.
 In the case of the $f_0(1710)$, as well as in the
$\sigma$ meson case to be discussed in the next section, the
decay of the resonance into a pair of neutral pions is present. A
pair of neutral pions can also be produced in a continuous two
photon fusion process. The rates for all the reactions discussed
in this work can be modified if both neutral pions, no matter if
they come from a resonance or a continuos process, are
misidentified with photons. This accidental background can be
easily isolated measuring its invariant mass distribution, and
making a cut that discriminates a single photon coming from the
processes that we are studying, from one that produced two
neutral mesons (mostly pions) subsequently decaying into two
photons. For example, in the case of the sigma meson (see next
section) each one of the neutral pions from its much large
hadronic decay should be misidentified. These pions would produce
pairs of photons with a large opening angle $\phi$, where
$\cos{(\phi /2)} = \sqrt{1 - 4m_{\pi}^{2}/m_{\sigma}^2}$.
\small{
\begin{table}[hb]
\begin{center}
\begin{tabular}{l c }
 particle & events/year \\
\hline
$\eta$  & $7.44 \times 10^{5}$  \\
$\eta^\prime$ & $2.67 \times 10^{4}$ \\
$f_0(1710)$ & $42$  \\
\end{tabular}
\vskip 0.3cm \caption{Number of events/year above background for
the $\eta$, $\eta^\prime$ and  $f_0(1710)$ resonances. }
\end{center}
\label{tab6}
\end{table}
}
%
However, the calorimeters already in use in many experiments are
able to distinguish between these two and single photon events
with high efficiency (see, for instance, Ref.\cite{alde}).

\section{Can we observe the $\sigma $ meson in peripheral collisions?}

The possible existence of light scalar mesons (with masses less
than about 1 GeV) has been a controversial subject for roughly
forty years. There are two aspects: the extraction of the scalar
properties from experiment and their underlying quark
substructure. Because the $J=0$ channels may contain strong
competing contributions, such resonances may not necessarily
dominate their amplitudes and could be hard to ``observe". In
such an instance their verification would be linked to the model
used to describe them.

Part of the motivation to study the two-photon final states in
peripheral heavy-ion collisions was exactly to verify if we can
observe such scalar mesons in its $\gamma\gamma$ decay. Although
this decay mode is quite rare, it has the advantage of not being
contaminated by the strong interaction of the hadronic final
states. In particular, it may allow to investigate the possible
existence of the sigma meson. This meson is expected to have a
mass between 400-1200 MeV and decay width between 300-500 MeV,
decaying predominantly into two pions. Of course, another decay
channel is into two photons, with the background discussed in
Section III.

Recently the E791 Collaboration at Fermilab found a strong
experimental evidence for a light and broad scalar resonance,
that is, the sigma, in the $D^+ \rightarrow \pi ^- \pi ^+ \pi ^+$
decay \cite{e791}. The resonant amplitudes present in this decay
were analyzed using the relativistic Breit-Wigner function
 given by
\begin{eqnarray}
BW = \frac{1}{m^2 - m^2_0 + i m_0 \Gamma(m)},
 \nonumber
\end{eqnarray}

\noindent
 with
\begin{eqnarray}
\Gamma(m) = \Gamma_0 \frac{m_0}{m} \left( \frac{p^*}{p^*_0}
\right) ^{2J + 1} \frac{^{J}F^2(p^*)}{^{J}F^2(p^*_0)},
 \nonumber
\end{eqnarray}

\noindent
 where $m$ is the invariant mass of the two photons forming a
 spin-J resonance. The functions $^JF$ are the Blatt-Weisskopf
 damping factors \cite{blatt}: $^0F = 1$ for spin 0 particles,
 $^1F = 1/\sqrt{1 + (rp^*)^2}$ for spin 1 and $^2F =  1/\sqrt{9 +
 3 (rp^*)^2 + (rp^*)^4}$ for spin 2. The parameter $r$ is the
 radius of the resonance ($\approx 3$ fm) \cite{argus} and $p^* =
 p^*(m)$ the momentum of decay particles at mass $m$, measured in
 the resonance rest frame, $p^*_0 = p^*_0(m)$, where $m_0$ is the
 resonance mass. The Dalitz-plot of the decay can hardly be
 fitted without a $0^{++}$ ($\sigma$) resonance.
The values of mass and total decay width found by the
collaboration with this procedure are $478 ^{+24}_{-23} \pm 17$
MeV and $324^{+42}_{-40} \pm 21$ MeV, respectively.

We will discuss if this resonance can be found in peripheral
heavy-ion collisions through the subprocess $\gamma \gamma
\rightarrow \sigma \rightarrow \gamma \gamma $. It is important
to note that all the values related to the $\sigma$, like mass or
partial widths, that can be found in the literature are very
different and model dependent. In particular, we find the result
of the E791 experiment very compelling and among all the
possibilities we will restrict ourselves to their range of mass
and total decay width, while we vary the partial width into two
photons.

\begin{figure}[htb]
\epsfxsize=.45\textwidth
\begin{center}
\leavevmode \epsfbox{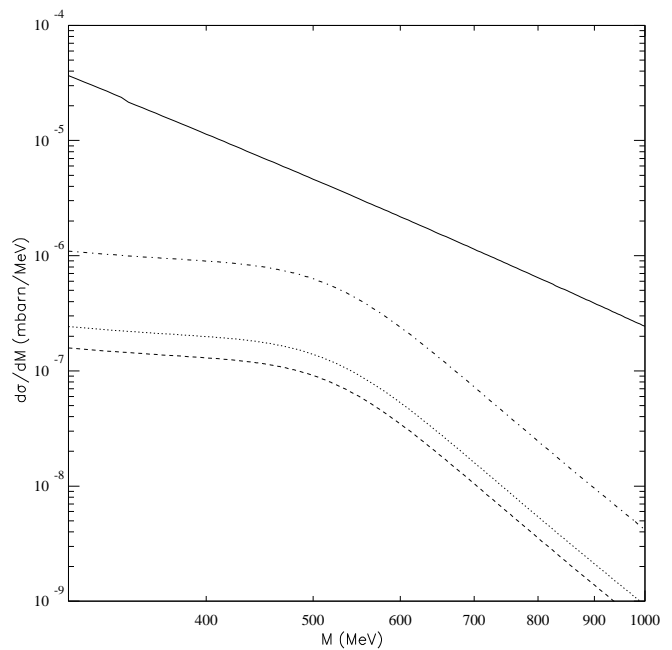}
\end{center}
\vskip -1.6cm
 \caption{Invariant mass distribution of photon production. The solid curve is
 the box diagram, the dashed curve is for the process
 $\gamma \gamma \rightarrow \sigma \rightarrow \gamma \gamma $
 when the $\Gamma _{\gamma \gamma} = 10 $ KeV, the dotted one is for
 $\Gamma _{\gamma \gamma} = 4.7$ KeV and the dotted-dashed one is for
 $\Gamma _{\gamma \gamma} = 3.8$ KeV. The angular cut is equal to
 $-0.5 < \cos \theta < 0.5$ in all curves.  }
  \label{caixa_sigma}
\end{figure}

For the $\sigma$ decay width into two photons we assume the
values obtained by Pennington and Boglione, $3.8 \pm 1.5$ KeV and
$4.7 \pm 1.5$ KeV \cite{pennington}, and the value of $10 \pm 6$
KeV \cite{courau}. The results obtained are in
Fig.(\ref{caixa_sigma}), the dashed curve is for $\Gamma _{\gamma
\gamma} = 3.8$ KeV, the dotted one is for $\Gamma _{\gamma
\gamma} = 4.7 $ KeV and the dotted-dashed one is for $\Gamma
_{\gamma \gamma} = 10$ KeV. In all cases we used the mass and
total decay width obtained by the E791 Collaboration, $m_\sigma =
478$ MeV and $\Gamma_{total} = 324$ MeV. In the same figure it can
be seen the curve due to the box diagram, In all cases we used
the angular cut $ -0.5 < \cos \theta <0.5$.

It has been verified in the case of $\pi \, \pi$ scattering that
the use of a constant total width in the $\sigma$ resonance shape
is not a good approximation\cite{schechter}. In our case we will
discuss the $\gamma \gamma \rightarrow \gamma \gamma$ process
above the two pions threshold where the peculiarities of the broad
resonance, basically due to the $\sigma$ decay into pions, are not
so important. Of course, another reason to stay above $300$ MeV
is that we are also far from the pion contribution to $\gamma
\gamma$ scattering. In any case we also computed the cross
section with a energy dependent total width $\Gamma(m) \simeq
\Gamma_0 \left( {p^*}/{p^*_0} \right) ^{2J + 1}$, which, as shown
by Jackson many years ago\cite{jackson}, is more appropriate for
a quite broad resonance. The net effect is a slight distortion of
the cross section shape with a small increase of the total cross
section. Since this one is a negligible effect compared to the
one that we will present in the sequence we do not shall consider
it again.

Up to now we assumed in this analysis only the Breit-Wigner
profile for the $\sigma$ contribution and it is interesting to
discuss the full process (including the $t$ and $u$ channels)
within a given model to verify how good is our approximation.

To compute the amplitudes for $\gamma \gamma \rightarrow \gamma
\gamma$ including the presence of a scalar resonance we can
assume the interaction $g_s F^{\mu \nu} F_{\mu \nu} \Phi_s $,
where $g_s$ is the coupling of the photons to the scalar field
$\Phi_s$. These amplitudes are
\begin{eqnarray}
M_{++++} &=& - \frac{2 \pi}{\alpha^2} F(\lambda r_t),
 \nonumber \\
M_{+-+-} &=& - \frac{2 \pi}{\alpha^2} F(\lambda t_t),
 \nonumber \\
M_{+--+} &=& - \frac{2 \pi}{\alpha^2} F(\lambda s_t),
 \nonumber \\
 M_{+++-} &=& 0,
  \nonumber \\
 M_{++--} &=& - \frac{2 \pi}{\alpha ^2} \{ F(\lambda s_t) + F(\lambda
 t_t) + F(\lambda r_t) \} .
\label{escalar}
\end{eqnarray}

Note that this interaction violates unitarity and is appropriate
only at energies up to the $\sigma$ mass.

\begin{figure}[htb]
\epsfxsize=.45\textwidth
\begin{center}
\leavevmode \epsfbox{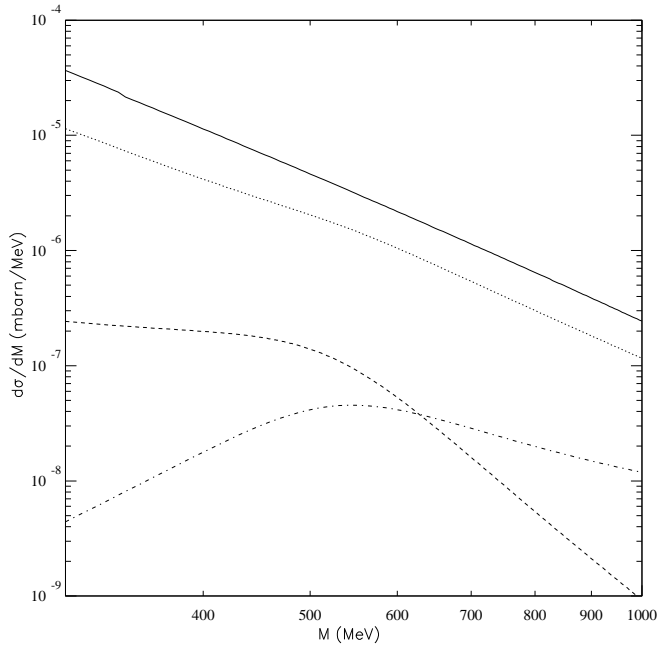}
\end{center}
\vskip -1.6cm
 \caption{ Invariant mass distribution of photon pair production.
 The solid curve is due to box diagram only, the dashed one is due to the process
 $\gamma \gamma \rightarrow \sigma \rightarrow \gamma \gamma$ in the
 Breit-Wigner approximation, the dash-dotted is the scalar contribution of
 Eq.(\ref{escalar}), and the dotted one is due to the total process,
 computed with box amplitudes plus the scalar boson amplitudes of
 Eq.(\ref{escalar}). In all cases $\Gamma_{\gamma\gamma} = 4.7$ KeV,
 and the angular cut is equal to $-0.5 < \cos \theta < 0.5$ .  }
  \label{interferencia_escalar}
\end{figure}

In Fig.(\ref{interferencia_escalar}) we show the full cross
section for the process $\gamma \gamma \rightarrow \gamma \gamma$
due to the box diagram and the existence of the scalar resonance
$\sigma$ with amplitudes given by Eq.(\ref{escalar}), computed
with $\Gamma_{\gamma\gamma} = 4.7$ KeV. Contrarily to the
previous cases the $\sigma$  is a quite large ``resonance" and
due to its large decay width off resonance the interference is
not negligible (the argument of the previous section obviouly
fails). Moreover, the interference is destructive! In
Fig.(\ref{interferencia_escalar}) we can see the continuous curve
which is due only to the box diagram, the dashed one is the cross
section of a scalar resonance computed with a Breit-Wigner
profile, Eq.(\ref{dsigfoton}), the dash-dotted curve is due only
to the scalar contribution of Eq.(\ref{escalar}), and the dotted
one is obtained when we consider the total cross section for two
photon production in peripheral heavy ion collisions, i.e., the
box diagram amplitudes plus the helicity amplitudes listed in
Eq.(\ref{escalar}). The figure shows clearly that the above
mentioned interference is destructive.

There is an important point to comment about the above result.
The effective Lagrangian model used to compute the $\sigma$
contribution to the photon pair production above $M \approx 600$
MeV gives a larger contribution to the cross section than the
Breit-Wigner one. We verified that the $t$ and $u$ channels
produce small variations in the result obtained when we consider
only the $s$ channel, but this one already grows as any
nonunitary amplitude calculated in this approach, and it is
unclear for us how far we can trust in this model. At small
masses the cross sections also differ in their behavior with $s$.

There are possible improvements that we may think of towards a
more reliable computation of the $\sigma$ meson exchange
amplitudes, but all of them are going to be plagued by
uncertainties as we remarked in the beginning of the section.
Therefore, we adopt the pessimistic view that the Breit-Wigner
result is just a lower bound for our process, because effective
lagrangians tend to increase the contribution at large masses
where the background vanishes rapidly, increasing the
signal/back\-ground ratio. The Breit-Wigner profile does the
opposite. The physical result is probably between both approaches.
From the experimental point of view we would say that the reaction
$\gamma \gamma \rightarrow \gamma \gamma$ has to be observed and
any deviation from the continuum process must be carefully
modeled until a final understanding comes out, with the advantage
that the final state is not strongly interacting. Note that in
this modelling the $\eta$ meson will contribute to $\gamma \gamma
\rightarrow \gamma \gamma$ in a small region of momentum, even so
it has to be subtracted in order to extract the complete $\sigma$
signal.

\vskip .7cm

\begin{figure}[htb]
\epsfxsize=.45\textwidth
\begin{center}
\leavevmode \epsfbox{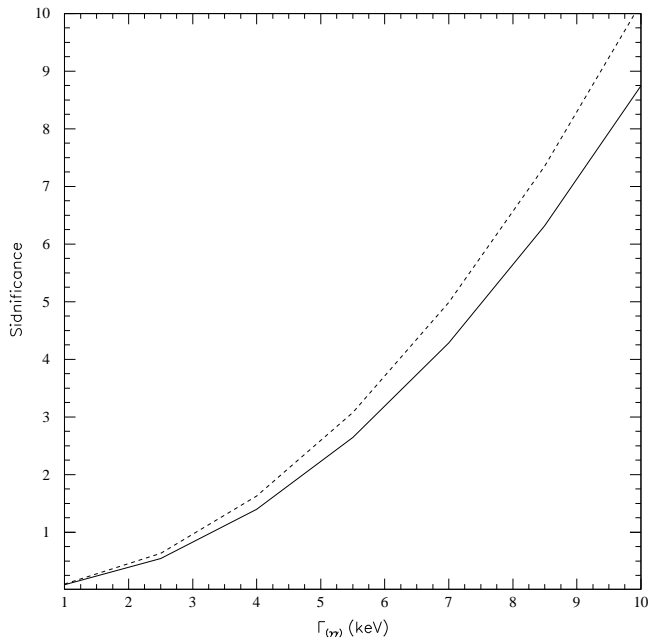}
\end{center}
\vskip -1.6cm
 \caption{ Significance as a function of decay width into two photons,
 $\Gamma _{\gamma \gamma}$, for a sigma meson with mass equal to 478 MeV and
 total decay width of 324 MeV. The solid curve was obtained integrating
 the cross sections in the interval $438<M<519$ MeV, the dashed one in
 the interval $300<M<800$ MeV.}
  \label{significancia}
\end{figure}

We changed the values of the $\sigma$ mass and total width around
the central ones reported by the E791 Collaboration. We do not
observed large variations in our result, but noticed that it is
quite sensitive to variations of the partial decay width into
photons. It is interesting to look at the values of the
significance which is written as $ {\cal
L}\sigma_{signal}/\sqrt{{\cal L}\sigma_{back}} $ and
characterizes the statistical deviation of the number of the
observed events from the predicted background. The significance
as a function of the two photons decay width of the sigma meson,
with mass equal to 478 MeV and total decay width of 324 MeV, is
shown in Fig.(\ref{significancia}), were we used a luminosity of
${\cal L} = 2.0 \times 10^{ 26}$ cm$^{-2}$s$^{-1}$ at RHIC and
assumed one year of operation. The significance is above
$2\sigma$ $95\%$ confidence level limit for two photon decay
width greater than 4.7 KeV while for a $5\sigma$ discovery
criteria can be obtained with $\Gamma_{\gamma \gamma} > 7.5$ KeV.
If $\Gamma_{\gamma\gamma}$ turns out to be smaller the numbers of
Fig.(\ref{significancia}) have to be scaled appropriately.

\vskip .7cm

The numbers in Fig.(\ref{significancia}) were computed with the
signal given by the Breit-Wigner profile, and the background by
the pure box diagram. The solid curve was obtained integrating
the cross sections in the interval $438<M<519$ MeV, corresponding
to the interval of mass uncertainty, while the dashed curve
resulted from the integration in the interval $300<M<800$ MeV.
Note that there is no reason, a priori, to restrict the
measurement to a small bin of energy. This choice will depend
heavily on the experimental conditions. When the signal is
extracted from Eq.(\ref{escalar}) and the cross sections are
integrated in the larger interval ($300-800$ MeV), we obtain
values for the significance more than one order of magnitude
above the curves of Fig.(\ref{significancia}), while the
significance is below the curves of Fig.(\ref{significancia}) up
to a factor $10$ when the cross sections are integrated in the
smaller interval of energy. The Breit-Wigner profile probably
gives an average of the results that could be obtained within
different model calculations. Therefore, for values of
$\Gamma_{\gamma\gamma}$ already quoted in the literature the
sigma meson has a chance to be seen in its two photon decay mode.
The discovery limits discussed above refer only to a statistical
evaluation. Our work shows the importance of the complete
simulation of the signal and background including an analysis of
possible systematic errors that may decrease the significance.

\section{Conclusions}

Peripheral collisions at relativistic heavy ion colliders provide
an arena for interesting studies of hadronic physics. One of the
possibilities is the observation of light hadronic resonances,
which will appear quite similarly to the two-photon hadronic
physics at $e^+e^-$ machines with the advantage of a huge photon
luminosity peaked at small energies~\cite{baur}. Due to this large
photon luminosity it will become possible to discover resonances
that couple weakly to the photons~\cite{natale}.

In this work we focus the attention at the peripheral reaction $Z
Z \rightarrow Z Z \gamma \gamma$. This process is important
because it may allow for the first time the observation of the
continuous subprocess $\gamma \gamma \rightarrow \gamma \gamma$
in a complete collider physics enviroment. This possibility only
arises due to enormous amount of photons carried by the ions at
the RHIC energies.

The continuous subprocess is described by a fermionic box diagram
calculated many years ago. We computed the peripheral heavy ion
production of a pair of photons, verifying which are the most
important contributions to the loop, which turned out to be the
electron at the energies that we are working, and stablished cuts
that not only ensure that the process is peripheral as well as
eliminate most of the background. After the cut is imposed we
still have thousands of photons pairs assuming 100$\%$ efficiency
in tagging the ions and detecting the photons.

The continuos $\gamma \gamma \rightarrow \gamma \gamma$ subprocess
has an interesting interplay with the one resulting from the
exchange of a resonance. We discuss the resonance production and
decay into a photon pair. This is a nice interaction to observe
because it involves only the electromagnetic couplings of the
resonance. Therefore, we may say that it is a clean signal of
resonances made of quarks (or gluons) and its measurement is
important because it complements the information obtained through
the observation of purely hadronic decays. It may also unravel
the possible amount of mixing in some glueball candidates
\cite{close}. We discuss the interference between these process
and compute the number of events for some specific cases.

The possibility of observing resonances that couple weakly to the
photons is exemplified with the $\sigma$ meson case. This meson,
whose existence has been for many years contradictory, gives a
small signal in the reaction $\gamma \gamma \rightarrow \sigma
\rightarrow \gamma \gamma$. However, its effects may be seen
after one year of data acquisition, providing some clue about
this elusive resonance. Using values of mass, total and partial
widths currently assumed in the literature, we compute the full
cross section within a specific model and discuss the
significance of the events. Our work shows the importance of the
complete simulation of the signal and background of these
processes including an analysis of possible systematic errors,
indicating that two photon final states in peripheral collisions
can be observed and may provide a large amount of information
about the electromagnetic coupling of hadrons.

\section*{Acknowledgments}

We would like to thank I.~Bediaga, C.~Dib, R.~Rosenfeld and
A.~Zimerman for many valuable discussions. One of us (CGR) thanks
M. de M. Leite for encouragement throughout this work and A. C.
Aguilar for useful discussions. This research was supported by
the Conselho Nacional de Desenvolvimento Cient\'{\i}fico e
Tecnol\'ogico (CNPq) (AAN), by Fundac\~ao de Amparo \`a Pesquisa
do Estado de S\~ao Paulo (FAPESP) (CGR,JPVC,AAN) and by Programa
de Apoio a N\'ucleos de Excel\^encia (PRONEX).


\begin {thebibliography}{99}

\bibitem{baur} C. A. Bertulani and G. Baur, {\it Phys. Rep.}
{\bf 163} 299 (1988); G.~Baur, J.\ Phys.\ {\bf G24}, 1657 (1998);
S. Klein and E. Scannapieco, hep-ph/9706358 (LBNL-40457); J.
Nystrand and S. Klein, hep-ex/9811997 (LBNL-42524); C. A.
Bertulani, nucl-th/0011065, nucl-th/0104059; J. Rau, b. M\"uller,
W. Greiner and G. Soff, {\it J. Phys. G: Nucl. Part. Phys.} {\bf
16} (1990) 211; M. Vidovi\'c, M. Greiner, C. Best and G. Soff,
{\it Phys. Rev.} {\bf C47} (1993) 2308.

\bibitem{gbaur} G. Baur, Proc. SBPF Intern. Workshop on
Relativistic Aspects of Nuclear Physics (Rio de Janeiro, 1989),
eds. T. Kodama et al. (World Scientific, Singapore, 1990), p. 127.

\bibitem{cahn} R.~N.~Cahn and J.~D.~Jackson,
Phys.\ Rev.\ {\bf D42}, 3690 (1990).

\bibitem{bertulani} G. Baur and C. A. Bertulani, {\it Nucl. Phys.}
{\bf A505} 835 (1989).

\bibitem{natale} A. A. Natale, {\it Phys.
Lett.} {\bf B362} 177 (1995); {\it Mod. Phys. Lett.} {\bf A9}
2075 (1994).

\bibitem{close} F. E. Close and A. Kirk, hep-ph/0103173
(OUTP-0114P).

\bibitem{e791} E791 Collaboration, E. M. Aitala et al., {\it Phys.
Rev. Lett.} {\bf 86} 770 (2001).

\bibitem{klein} J. Nystrand and S. Klein (LBNL-41111)
(Two-Photons Physics in Nucleus-Nucleus Collisions at RHIC)
nucl-ex/98110072.

\bibitem{engel} R. Engel et al., {\it Z. Phys.} {\bf C74} 687
(1997).

\bibitem{eu} C. R. Rold\~ao and A. A. Natale, {\it Phys. Rev.}
{\bf C61} (2000) 064907.

\bibitem{karplus} R. Karplus and M. Neuman, {\it Phys. Rev.} {\bf 83} (1951) 776.

\bibitem{tollis2} B. De Tollis, {\it Nuovo Cimento} {\bf 32} (1964)
757; {\bf 35} (1965) 1182.

\bibitem{landau} V. B. Berestetskii et al., in {\it Quantum
Electrodynamics} vol. 4, second edition (1996) 571.

\bibitem{tollis1} V. Costantini et al., {\it Nuovo Cimento} {\bf 2A}
(1971) 733.

\bibitem{spence} K. Mitchell, {\it Phil. Mag.} {\bf 40} (1949)
351.

\bibitem{tollis3} B. De Tollis and G. Violini, {\it Nuovo Cimento} {\bf 41A} (1966) 12.

\bibitem{ora} V. N. Oraevskii, {\it Sov. Phys. JETP} {\bf 12} (1961) 730; M. Y. Han
and S. Hatsukade, {\it Nuovo Cimento} {\bf 21} (1961) 119.

\bibitem{pennington} M. Boglione and M. R. Pennington, {\it Eur.
Phys. J.} {\bf C9} (1999) 11.

\bibitem{pdg} D. E. Groom et al. (Particle Date Group), {\it Eur.
Phys. J.} {\bf C15} (2000) 1.

\bibitem{courau} A. Courau et al., {\it Nucl. Phys.} {\bf B271}
(1986) 1.

\bibitem{schechter} F. Sannino and J. Schechter, {\it Phys. Rev.} {\bf D52} (1995) 96;
M. Harada, F. Sannino and J. Schechter, {\it Phys. Rev.} {\bf
D54} (1996) 1991; {\it Phys. Rev. Lett.} {\bf 78} (1997) 1603.

\bibitem{jackson} J. D. Jackson, {\it Nuovo Cimento} {\bf 34} (1964) 1644.

\bibitem{novaes} S. F. Novaes, {\it Phys. Rev.} {\bf D27} (1983) 2211.

\bibitem{alde} D. Alde et al., {\it Z. Phys.} {\bf C36} (1987) 603.

\bibitem{blatt} J. M. Blatt and V. F. Weisskopf, in {\it
Theoretical Nuclear Physics}, John Wiley $\&$ Sons, New York,
1952.

\bibitem{argus} ARGUS Collaboration, H. Albrecht et al., {\it
Phys. Lett.} {\bf B 308} (1993) 435.

\bibitem{close2} F. E. Close, G. R. Farrar and Z. Li, {\it Phys. Rev.} {\bf D55}
(1997) 5749.

\bibitem{aleph} ALEPH Collab., {\it Phys. Lett.} {\bf B422} (2000)
189.

\end {thebibliography}
\end{document}